\documentclass[10pt,twocolumn]{article}
\usepackage{enumitem}
\usepackage[margin=1.0in]{geometry}
\setlist{noitemsep}

\begin{document}

\title{Code Review For and By Scientists}
\author{
  Marian Petre \\ Open University \\ m.petre@open.ac.uk
  \and
  Greg Wilson \\ Mozilla Science Lab \\ greg@mozillafoundation.org
}

\maketitle

\begin{abstract}

We describe two pilot studies of code review by and for scientists.
Our principal findings are that scientists are enthusiastic,
but need to be shown code review in action,
and that just-in-time review of small code changes is more likely to succeed
than large-scale end-of-work reviews.

\end{abstract}

\section{Introduction}

Since Fagan's work in the 1970s \cite{b:fagan1976,b:fagan1986},
dozens of studies have shown that code review is
the most effective way to find bugs \cite{b:cohen2010,b:bacchelli2013}.
Ironically,
given that Fagan and others were inspired by academic peer review,
code review is still rare in scientific software development.
This is partly because most authors don't publish their code,
and hence have little incentive to improve its quality,
but also a case of the blind (not) leading the blind.

In 2013--14,
we ran two pilot studies to explore the benefits of code review for typical scientist-coders,
and how to transfer the skill to them.
Our focus was not the small minority of scientists who already use good software engineering practices \cite{b:hannay2009},
but on the large majority who have never encountered them.
This paper describes our methodology and findings,
and makes recommendations for other groups who wish to help scientists adopt code review.

\subsubsection*{Acknowledgments}

Our thanks to the scientists and programmers who took part in this study,
and to PLOS, the Mozilla Science Lab, and the Sloan Foundation for their support.

\section{Post-Hoc Reviews}

Our first study,
done in conjunction with the Public Library of Science (PLOS),
ran from August to October 2013.
Professional developers working at Mozilla
who routinely use asynchronous code review in their work
reviewed samples of code
taken from papers published in \emph{PLOS Computational Biology} in the preceding 18 months.
Their reviews were shared with the scientists;
we then conducted semi-structured interviews \cite{b:rosenthal2007,b:bryman2008} with 11 developers in late August and early September,
and with 4 authors whose code had been reviewed in late September and early October,
to determine
whether non-scientists could usefully review typical scientific software,
whether those reviews were intelligible and useful to the scientists,
and whether the participants felt the reviews were valuable.
Four significant findings emerged.

\noindent \textbf{1) Scientists' starting points varied widely.}
Scientists' prior knowledge and practices varied widely.
Some used engineering practices and tools such as version control some of the time,
but documentation of the form developers expected was rare,
and practices like continuous integration were unknown or considered ``very unusual''.

In particular,
most scientists had never taken part in code review.
Some of the code authors we interviewed work in (or lead) teams in which they read each other's code,
but most participants had never, or only rarely, discussed their code with others.
Instead, they discussed the results the code produced.

\noindent \textbf{2) Developers felt limited.}
Most developers repeatedly noted
the lack of documentation, commenting, tests, and example data sets,
e.g.,
``Not having the project build is a big problem; I can't verify that the code is correct.''
Equally,
most reviewers were frustrated at not being able to run the code as the first step in review,
e.g.,
``That's the easiest way to see if it works at all,''
and
``It's a way to validate the intentions of the author.''
Several commented that, as a result, they had to make some assumptions in the review.

Few reviewers read in detail the paper with which the code was associated,
usually because they lacked the domain expertise to do so.
They could deduce what the code did and assess it in terms of its ability to perform that operation efficiently,
but they could not tell if the code fulfilled its intended scientific role.
That,
combined with their isolation from the scientists,
left developers feeling that they were doing ``drive-by'' reviews.

\noindent \textbf{3) Standards and expectations were very different.}
Many reviewers were struck by how scientists' code differed from theirs,
particularly in lacking commenting and explanation in the code.
Several suggested that scientists appear to have ``less concern for maintainability and readability''
and that the code was ``not written for others to use''.
Some also pointed out naive lack of complexity or abstraction,
redundancy in the code,
inconsistencies in or ignorance of standards in formatting,
and unhelpful naming.

On their side,
the scientists all aspired to create readable and re-usable code,
but many noted that their code doesn't respect the common etiquette of open source:
for example,
they often don't make an effort to package their code for re-use by others.
One scientist commented that:
``[It's] not important to have something that's exact---only when we publish''
and reiterated the low status of code in their science:
``In the business of science, all that matters is the figures. The quality of the code is just not on the critical path.''
This echoes findings from other empirical research \cite{b:segal2005,b:segal2007}.

\noindent \textbf{4) Both sides wanted more contact.}
All but one of the reviewers remarked that they miss the social context they normally associate with code reviews:
working relationships, understanding goals and priorities, and trust.
They would have preferred some form of dialog,
not least to provide them with the code context,
and to establish appropriate expectations for both reviewers and authors
(e.g., ``knowing what kind of feedback the author wants'').
A typical remark was,
``It would have been easier if we were allowed to contact the scientist just to get a feel for his mindset.''
Most referred to dialogs that are normally part of their code review experience,
whether spoken or on-line:
``Discussion catches details that get lost otherwise''
such as ``tiny changes in numerical interpretation that are important''.
The reviewers also pointed out that the dialog has value for the reviewer,
as well as for the author:
``When there is a dialog, you end up learning a lot yourself.''

\section{Transferring the Skill}

Despite these frustrations,
participants in this study were positive about it.
We therefore set up a second study
that paired experienced scientific programmers with small groups of less-experienced ones
to explore ways of transferring the practice itself.
This second study also aimed to address the fact that
scientists and developers alike wanted reviews during development,
when they could act on the feedback,
rather than a ``drive-by'' review after submission.

Ten groups ranging in size from a couple of people to half a dozen initially signed up to take part in this study.
In interviews and early discussions,
they identified four main reasons why they think they ought to do code review:

\noindent \textbf{1) Rigor:}
Scientists want to get the right answer and be able to reproduce their work later.
(Most recognized that correctness and reproducibility aren't the same thing.)

\noindent \textbf{2) Reusability:}
Scientists are very aware that their understanding of code dissipates over time
and that this is a large hidden cost.
Equally, they suspect that they spend a lot of time reinventing wheels.
They may not know how code review will help with that, but they hope that it will.

\noindent \textbf{3) Collaboration:}
Many scientists hoped that they could use code review
as an excuse for conversations about code with their colleagues---conversations
that simply don't happen right now.
Some believed that review would foster better testing,
encourage scientists to produce code that is easier to understand and share,
and make team members more aware of each other's research.

\noindent \textbf{4) Knowledge transfer:}
All participants felt that there must be a better way to build programs than what they're doing right now.
Taking part in this study was,
for them,
a way to get mentoring about programming in general from someone who could answer questions they didn't know to ask.

One thing we \emph{didn't} hear was people saying that
they wanted to learn about the science embodied in the code being reviewed.
Rightly or wrongly,
scientists seem to feel that ``what it does'' can be learned in other ways.

Each team was paired with a mentor:
another scientist-developer with experience of code review,
from a different institution but usually in the same or a cognate discipline,
who would introduce the team to code review processes and good practices.
Teams and mentors collaborated during March--June 2014.
Each mentor decided how to introduce and conduct reviews with the team,
informed by the mentor's own experience and discussion with the team.
The scientists identified which of their code to review, in consultation with the mentor.
We followed teams' interactions on email (via shared mailing lists) and within their repositories,
attended meetings when possible (using video or telephone conferencing),
and interviewed participants periodically through the study about their goals, experiences, and impressions.

Our biggest finding is how much \emph{perspective matters to collaboration}.
Over time,
professional software developers
(and scientists whose primary role has become programming)
integrate skills and knowledge,
building up a standard vocabulary of terms, tools, and conventions,
as well as a practical repertoire that it's easy to take for granted.

Most scientists don't (yet) have such an overview:
they are focused almost entirely on the purpose the code serves rather than the code itself.
They often lack the integrated understanding and skill that comes with experience,
which meant it was often very useful for mentors to ``state the obvious''.

How to make the transition from the first perspective to the second is usually not clear to working scientists.
``Stuff that seemed like overkill makes sense now...
A lightbulb finally went off:
I understand how other people look at my code and how to make it work.''
And what basics to articulate---where to look and how to look---is often not clear to experienced developers.
If the two are going to collaborate,
then at some point the different perspectives must be acknowledged and bridged.
Social mechanisms like courtesy, deference, and avoiding embarrassment tend to obscure this,
so it's crucial to articulate collaborators' perspectives explicitly for both sides,
and equally to help establish appropriate expectations explicitly.

This observation isn't new:
as Segal discussed in her study of industrial software engineers collaborating with scientists \cite{b:segal2005},
``[Programmers] demand an up-front articulation of requirements,
whereas the scientists had experience, and hence expectations, of emergent requirements.
The project documentation does not suffice to construct a shared understanding.''

For every mentor who says,
``I wish I knew more clearly what the scientists are expecting,''
there's a scientist who says,
``I wish I knew what it's reasonable to expect.''
One concrete step we could take is to show scientists what it actually looks like to do a code review:
not just the end result, but the process itself.

The most successful groups---those that completed the pilot study,
made code review part of their daily practice,
and considered it beneficial---both saw code review done and did it themselves.
Their mentors walked through an initial code review;
articulated what they look for in the code, what they see, and how they interpret it;
and made concrete suggestions about first steps that clarified scientists' expectations.
But participants (both scientists and mentors) from all of the successul teams
perceived the real benefits only after they undertook the process themselves,
engaging in code review for each other,
and discussing those reviews with others.

Another observation was that \emph{there are several models of peer review},
each with its own emphasis.
In \emph{asynchronous review},
which is the predominant model in open-source development,
comments and revisions are mediated by the version control system.
\emph{Offline review with synchronous presentation}
uses the tools available in the version control system to present the code and supporting material for review,
and to capture the reviewer's comments,
but the reviewer and development team meet (usually virtually) to talk through the code and the review
in order to facilitate a freer interaction,
especially in terms of questions and clarifications.
And finally,
\emph{synchronous review}
involves walking and talking through the code together,
whether one-to-one, as a group, or with one person leading.

There are other axes of variation:
is review one-to-one, one-to-many, or done in a group?
Does it use advance preparation or is done on the day?
Is there a leader?
These choices should depend on whether the goal is
fixing code, learning from code, or learning from other developers:
while offline preparation helps developers form independent opinions,
synchronous activity helps develop rapport and trust.

The groups that were most successful in the second study
took the time for a synchronous conversation to introduce themselves and to discuss their goals and expectations
(often in addition to introductory emails).
Their mentors set out a review process and timetable at the outset,
based on their experience.
This usually involved a synchronous conversation of fixed length at regular intervals (3-4 weeks)
with interim goals and tasks agreed during each of those conversations.

Finally,
\emph{do not underestimate returns from small investments}.
None of the mentors expected scientists to overhaul complete code bases.
The advice from one mentor was cogent:
if you check the docstring and write a test every time you touch a method,
the code improvements will accumulate over time with minimal effort.
This insight was echoed in different ways throughout the study.
Adopting good practices and applying them incrementally,
doing one thing at a time as it is needed,
will achieve changes faster than you might think.

\section{Conclusions}

All of the mentors who engaged in the pilot study did so because,
in their experience,
the benefits of code review to science are profound in terms of improving reproducibility,
promoting re-usability,
disseminating best practices,
and thereby improving efficiency and avoiding error.
In other words,
they believe code review improves their \emph{science}
as well as the code itself.
Moreover,
they believe that the cost of introducing code review is quickly recovered,
and that good practices can be introduced incrementally in ways that are not disruptive or expensive.

The scientists engaged in the pilot study because they wanted to write more accurate, efficient, and re-usable code,
and wanted support in learning how to do so.
They too perceived that improvements to code development pay off in terms of their science,
and code review provides an otherwise scarce opportunity to ``talk code'' and learn better practices.
Many of the scientists reported benefits even during the study:
code improvements,
better use of facilities provided in development tools,
and better documentation practices spreading in the lab.  

Nevertheless,
many teams found it difficult to reserve the time to work with the mentor,
given lab priorities such as bidding for further funding and meeting publication deadlines.
And priorities were not the only obstacles:
identifying specific code for review (from a larger code base) was not always obvious,
the code was not always well-modularized and was rarely well-documented,
and teams often had to put code into a repository for the very first to make it available for review.
Some groups also needed to negotiate issues of confidentiality (for datasets as well as code).

Based on our study,
we believe the following should be priorities when scientists start doing code review:

\begin{enumerate}

\item Have a goal, a benefit in mind.

\item Start with a conversation:
articulate your goals and expectations, build rapport.

\item Choose the right pieces of code for the first reviews.
A good starting point is 3--4 pages long,
fairly self-contained,
and under active development,
so that small patches are coming in regularly.

\item Make your code available in a repository
with a typical data set
and an overview of how the code works.

\item Set up a schedule and commit a little time on a regular basis.

\item Understand that it's the code that's being reviewed (not you).
Tailor your comments to others on the same terms.

\item Make the process reciprocal:
be prepared to make as well as receive comments.

\end{enumerate}

We have learned a lot from these two studies,
but they are only the beginning.
As our work progresses,
we hope to:

\begin{enumerate}

\item
articulate a curriculum for scientists who would like to adopt code review,

\item
provide examples of people doing review and talking about their reasoning
(most likely in the form of short screencasts of reviews being done),

\item
develop heuristics to help people decide
which scientific software needs specialist scientific knowledge
and which doesn't,
and

\item
track what happens to code in the months \emph{after} review.

\end{enumerate}

If you would like to help with this work,
we would enjoy hearing from you.

{\small
\bibliographystyle{unsrt}
\bibliography{wssspe2014-code-review}

\begin{thebibliography}{1}

\bibitem{b:fagan1976}
M.~Fagan.
\newblock Design and code inspections to reduce errors in program development.
\newblock {\em IBM Systems Journal}, 15(3):182--211, 1976.

\bibitem{b:fagan1986}
M.~Fagan.
\newblock Advances in software inspections.
\newblock {\em IEEE Trans. Software Engineering}, 12(7):744--751, July 1986.

\bibitem{b:cohen2010}
J.~Cohen.
\newblock Modern code review.
\newblock In A.~Oram and G.~Wilson, editors, {\em Making Software}. O'Reilly,
  2010.

\bibitem{b:bacchelli2013}
Alberto Bacchelli and Christian Bird.
\newblock Expectations, outcomes, and challenges of modern code review.
\newblock In {\em Proc.\ International Conference on Software Engineering}, May
  2013.

\bibitem{b:hannay2009}
Jo~Erskine Hannay, Hans~Petter Langtangen, Carolyn MacLeod, Dietmar Pfahl,
  Janice Singer, and Greg Wilson.
\newblock How do scientists develop and use scientific software?
\newblock In {\em Second International Workshop on Software Engineering for
  Computational Science and Engineering (SECSE09)}, 2009.

\bibitem{b:rosenthal2007}
R.~Rosenthal and R.I. Rosnow.
\newblock {\em Essentials of Behavioural Research: Methods and Data Analysis}.
\newblock McGraw-Hill, 3 edition, 2007.

\bibitem{b:bryman2008}
A.~Bryman.
\newblock {\em Social Research Methods}.
\newblock Oxford University Press, 3 edition, 2008.

\bibitem{b:segal2005}
Judith Segal.
\newblock When software engineers met research scientists: A case study.
\newblock {\em Empirical Software Engineering}, 10(4):517--536, 2005.

\bibitem{b:segal2007}
Judith Segal.
\newblock Some problems of professional end user developers.
\newblock In {\em IEEE Symposium on Visual Languages and Human-Centric
  Computing}, pages 111--118, 2007.

\end{thebibliography}
}

\end{document}